\documentclass[prl,aps,twocolumn,epsf]{revtex4}

\def\y{\'{\i}}
\def\ni{\noindent}

\begin{document}
\title{Fluctuations of the Initial Conditions and the 
       Continuous Emission in Hydrodynamic Description 
       of Two-Pion Interferometry} 
\author{O. Socolowski Jr.$^1$, F. Grassi$^1$, 
        Y. Hama$^1$          
        and T. Kodama$^2$}
\affiliation{$^1$ Instituto de F\y sica, Universidade 
         de S\~ao Paulo, C.P. 66318, 05315-970 
         S\~ao Paulo-SP, Brazil \\
         $^2$ Instituto de F\y sica, Universidade 
         Federal do Rio de Janeiro, C.P. 68528, 
         21945-970 Rio de Janeiro-RJ , Brazil} 
\maketitle

\vspace*{-.8cm} 
\hspace*{1.3cm} 
\parbox{14.7cm}
{\hspace{.5cm}\small{
Within hydrodynamic approach, we study the Bose-Einstein 
correlation of identical pions by taking into account 
both event-by-event fluctuating initial conditions and 
continuous pion emission during the whole development of 
the hot and dense matter formed in high-energy 
collisions. Important deviations occur, compared to 
the usual hydro calculations with smooth initial 
conditions and a sudden freeze-out on a well defined 
hypersurface. Comparison with data at RHIC shows that, 
despite rather rough approximation we used here, this 
description can give account of the $m_T$ dependence of 
$R_L$ and $R_s$ and produces a significant improvement 
for $R_o$ with respect to the usual version. 

\medskip

\ni PACS numbers: 25.75.-q, 24.10.Nz, 24.60.-k
}} 

\bigskip
\vspace*{.05cm}

\ni {\it Introduction} -- When describing 
ultra-relativistic heavy-ion collisions in hydrodynamic 
approach\cite{landau}, a simple picture has been 
extensively adopted. It is usually considered that, 
after the initial very complicated interaction between 
two incident nuclei, at some early instant of time a 
local thermal equilibrium is attained. Such a state is 
usually described in terms of a set of {\bf highly 
symmetric} and {\bf smooth distributions of velocity} 
and {\bf thermodynamical quantities}. These are the 
initial conditions (IC) for the hydrodynamic equations, 
which must be complemented with reasonable equations of 
state (EoS). Then, as the thermalized matter expands, 
the system gradually cools down and, when the 
temperature reaches a certain freeze-out value 
$T_{fo}$, it {\bf suddenly} decouples. Every observed 
quantity is then computed on the hypersurface 
$T=T_{fo}$. For instance, the momentum distribution of 
the produced hadrons are obtained by using the 
Cooper-Frye integral\cite{CF} extended over this 
hypersurface. 

Though operationally simple, and actually useful for 
obtaining a nice comprehension of several aspects of 
the phenomena, such a scenery is clearly highly 
idealized when applied to finite-volume and 
finite-lifetime systems as those formed in high-energy 
heavy-ion collisions. In this letter, we examine 
modification of two ingredients of such a description, 
namely, {\it i)} effects of {\bf fluctuations in the 
IC}; and {\it ii)} consequences of {\bf continuous 
emission} (CE) of particles, regarding two-pion 
correlation. 

The identical-particle correlation, also known as HBT 
effect\cite{hbt1,hbt2} is a powerful tool for probing 
geometrical sizes of the space-time region from which 
they were emitted. If the source is static like a star, 
it is directly related to the spatial dimensions of the 
particle emission source. When applied to a dynamical 
source, however, several non-trivial effects 
appear\cite{hama,pratt}, reflecting its time evolution 
as happens in high-energy heavy-ion collisions. Being 
so, the inclusion of IC fluctuations and of the 
continuous emission may affect considerably the 
so-called HBT radii, because both of them modify in an 
essential way the particle emission zone in the 
space-time. 

The usual symmetric, smooth IC may be 
understood\hfilneg\ 

\newpage 
\vspace*{2.5cm} 

\ni as corresponding to the mean distributions of hydrodynamic 
variables averaged over several events. However, since 
our systems are not large enough, large fluctuations 
varying from event to event are expected. In previous 
publications\cite{ebe1,ebe2}, we showed that indeed the 
effects of these fluctuations on the observed 
quantities are sizeable and moreover in\cite{ebe2}, we 
compared the rapidity distributions of pions and 
$p-\bar p$ and showed that the average multiplicicity 
of pions decreases if we consider the event-by-event 
fluctuations, in comparison with the one given by 
the averaged IC\cite{hirano}. Concerning 
two-pion correlation, as the IC in the event-by-event 
base often show small high-density spots in the energy 
distribution, our expectation is that these spots 
manifest themselves at the end when particles are 
emitted, giving smaller HBT radii. 

As for the decoupling process, it has been 
proposed\cite{ghk,laszlo} an alternative picture where 
the emission occurs not only from the sharply defined 
freeze-out hypersurface, but continuously from the whole 
expanding volume of the system at different temperatures 
and different times. According to this picture, the 
large-transverse-momentum $(k_T)$ particles are mainly 
emitted at early times when the fluid is hot and mostly 
from its surface, whereas the small-$k_T$ components are 
emitted later when the fluid is cooler and from larger 
spatial domain. Mostly by using a simple scaling 
solution, we 
showed in the previous papers that this picture gives 
several nice results, namely, {\it i}) CE enhances the 
large-$m_T$ component of the heavy-particle 
($p,\Lambda,\Xi,\Omega,...$) spectra, {\it ii}) it gives 
a concave shape for the pion $m_T$ spectrum even without 
transverse expansion of the fluid\cite{ghk,laszlo}, 
{\it iii}) it can 
lead to the correct hyperon production ratios and 
spectrum shapes with conceptually reasonable choice of 
parameters\cite{ghk,ghks,gs1}, and {\it iv}) it 
reproduces the observed mass dependence of the slope 
parameter $T$ \cite{gs2}. Concerning HBT correlation, 
we showed\cite{hbt-cem}, within the same approximation, 
that whereas the so-called {\it side} radius is 
independent of the average $k_T\,$, the {\it out} radius 
decreases with $<k_T>$, because of the reason mentioned 
above. This behavior is expected to essentially remain 
in the general case we are going to discuss below and 
shown by data\cite{star}. 
\smallskip

\ni {\it Initial Conditions} -- In order to produce 
event-by-event fluctuating IC, we use the NeXus event 
generator\cite{nexus}, based on Gribov-Regge model. 
Given the incident nuclei and the incident energy, it 
produces the energy-momentum tensor distribution at the 
time $\sqrt{t^2-z^2}=1~$fm, in event-by-event basis. 
This, together with the baryon-number density 
distributions, constitute our fluctuating IC. The 
strangeness has not been introduced in the present 
calculations. As mentioned in the Introduction, we 
understand that the usual symmetric, smooth IC may be 
obtained from these by averaging over many events. In 
Fig. 1, we show an example of such an event for central 
Au+Au collision at 130 A GeV, compared with an average 
over 30 events. As can be seen, the energy-density 
distribution for a single event (left), at the 
mid-rapidity plane, presents several blobs of 
high-density matter, whereas in the averaged IC (right) 
the distribution is smoothed out, even though the number 
of events is only 30. In the results below, we show that 
the fingerprints of such high-density spots remain until 
the frezeout stage of the fluid, giving smaller HBT 
radii. 

\begin{figure}[tbh]
\vspace*{-1.cm} 
\begin{center}
\includegraphics{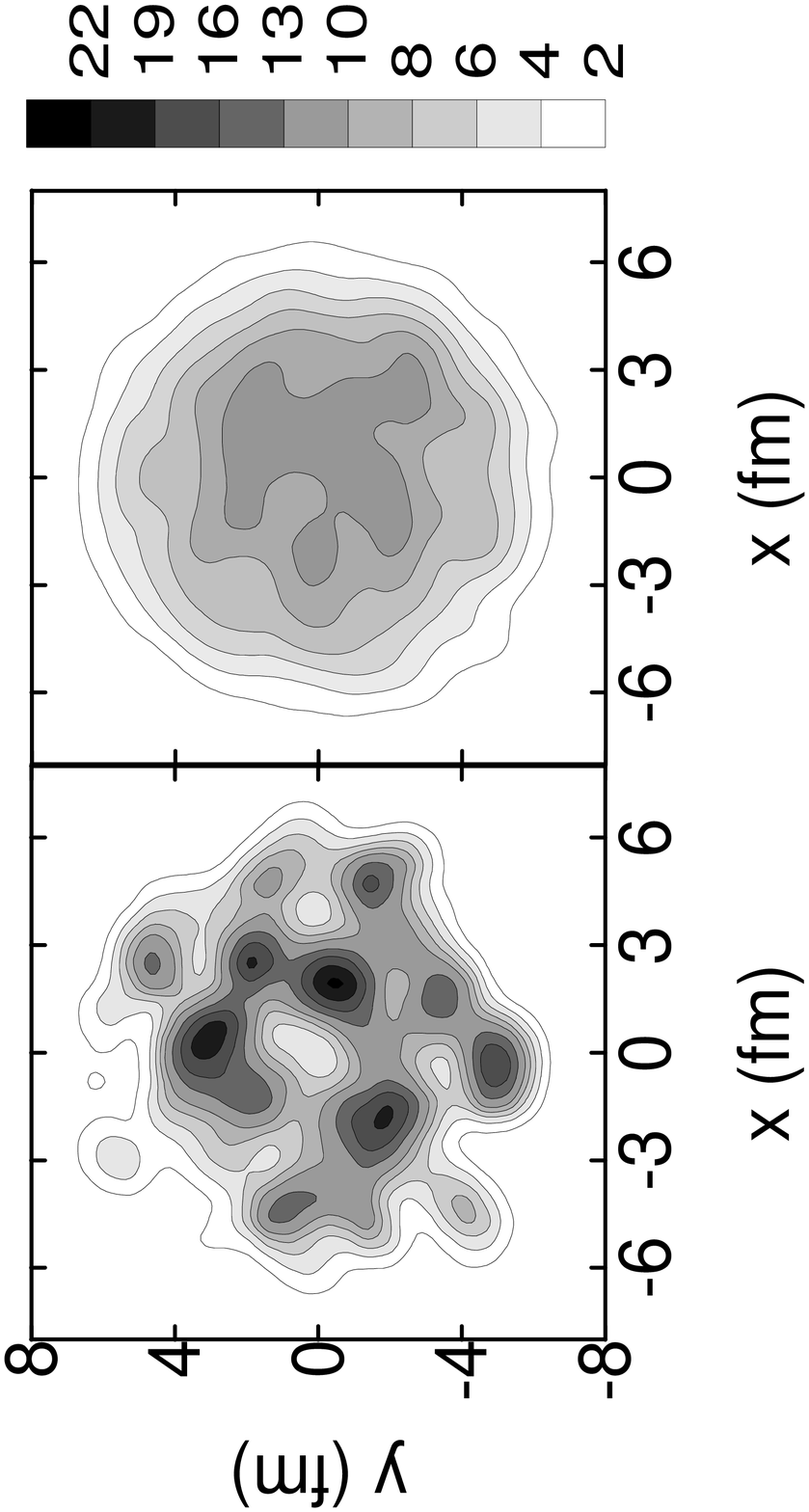}
\end{center}
\vspace{4.2cm} 
\caption{\small Examples of initial conditions for 
central Au+Au collisions given by NeXus at mid-rapidity 
plane. Energy density is plotted in units of GeV/fm$^3$. 
Left: one random event. Right: average over 30 random 
events (corresponding to the smooth initial conditions 
in the usual hydro approach).} 
\end{figure} 
\vspace*{-.2cm}

\ni {\it Hydrodynamic Equations} -- The resolution of 
the hydrodynamic equations deserves a special care, 
since our initial conditions do not have, in general, 
any symmetry nor they are smooth. We adopt the recently 
developed SPheRIO code\cite{spherio1}, 
based on the so-called 
smoothed-particle hydrodynamics (SPH), first used in 
astrophysics and which we have adapted for nuclear 
collisions$\,$\cite{spherio}, a method flexible enough, 
giving a desired precision. The peculiarity of SPH is 
the use of discrete Lagrangian coordinates attached to 
small volumes (``particles'') with some conserved 
quantities. Here, we take the entropy and the baryon 
number as such quantities. Then, the entropy density, 
for example, is parametrized as 
\begin{equation} 
 s({\bf x},t)=\frac{1}{\gamma}\sum_i \nu_i~W({\bf x}-
 {\bf x}_{\,i}(t);h)~,\label{s} 
\end{equation} 
where 
$W({\mathbf x}-{\mathbf x}_{\,i}(t);h)$ is a 
normalized kernel; 
{\bf x}$_i(t)$ is the $i$-th particle position, 
so the velocity is {\bf v}$_{\,i}=d{\bf x}_i/dt\,$;  
$h$~is the smoothing scale parameter, and we have 
\newpage 
\begin{equation} 
 S=\int\!d^3{\bf x}~\gamma s({\bf x},t)=\sum_i^N \nu_i~. 
\end{equation} 
Observe that, since an entropy $\nu_i$ is attached to 
the $i$-th SPH particle, and so is a baryon number, 
the total entropy $S$ and the baryon number $N$ are 
automatically conserved. By rewriting the usual 
energy-momentum conservation equation 
$\partial_\mu T^{\mu\nu}=0$, we get a set of coupled 
ordinary equations\cite{spherio} 
\begin{eqnarray} 
 \frac{d}{dt}\left(\nu_i\frac{P_i+\varepsilon_i}{s_i}
       \gamma_i{\bf v}_i\right)\hskip4.5cm&&\nonumber \\ 
 = -\sum_j\nu_i\nu_j\bigg[\frac{P_i}{\gamma^2_i{s_i}^2}
       +\frac{P_j}{\gamma^2_j{s_j}^2}\bigg]\, 
    {\bf\nabla}_i W({\bf x}_{\,i}-{\bf x}_{\,j};h)\,. &&  
\end{eqnarray} 
\ni {\it Equations of State} -- For equations of state, 
we consider ones with a first-order phase transition 
between QGP and a hadronic resonance gas, with baryon 
number conservation taken into account. In the QGP, we 
consider an ideal gas of massless quarks $(u,d,s)$ and 
gluons, with the bag pressure $B$ taken to give the 
critical temperature $T_c=160\,$MeV at zero chemical 
potential. In the resonance gas phase, we include all 
the resonances below 2.5 GeV, with the excluded volume 
effect taken into account. 
\vspace*{-.2cm}

\ni {\it Two-Pion Correlation} -- We assume that all 
pions are emitted from a chaotic source and neglect the 
resonance decays. It is argued\cite{csorgo} that, since 
resonance decays contribute to the correlations with 
very small $q$ values 
($q\lower.1cm\hbox{$\,\buildrel <\over\sim\,$}q_{min}\,$, 
where $q_{min}$ is the minimum measureable $q$), 
the experimentally determined HBT radii are essentially 
due to the direct pions. Then the correlation function 
is expressed in terms of the distribution function 
$f(x,k)$ as 
\vspace*{-.2cm} 
\begin{equation} 
C_2(q,K)=1+\frac{|I(q,K)|^2}{I(0,k_1)I(0,k_2)} \label{hbt}
\end{equation} 
where $K=(k_1+k_2)/2$ and $q=(k_1-k_2)$ and $k_i$ is 
the momentum of the $i$th pion. Usually 
\begin{equation} 
 I(q,K)\equiv\langle a_{k_1}^+a_{k_2}\rangle
 =\int_{T_{fo}}d\sigma_\mu K^\mu f(x,K)e^{iqx}. \label{I} 
\end{equation} 
In SPH representation, we write $I(q,K)$ as 
\begin{equation} 
I(q,K)=\sum_j 
\frac{\nu_j n_{j\mu} K^\mu}{s_j |n_{j\mu} u_j^\mu|}\;
{\mathrm{e}}^{iq_\mu x_j^\mu}f(u_{j\mu} K^\mu)\,,\label{corr_sph} 
\end{equation} 
where the sumation is over all the SPH particles. In the 
Cooper-Frye freezeout, these particles should be taken 
where they cross the hyper-surface $T=T_{fo}$ and 
$n_{j\mu}$ is the normal to this hyper-surface. 
\smallskip

\ni {\it Continuous Emission Model} -- In CEM\cite{ghk}, it is assumed that, at each space-time point $x^\mu $, 
each particle has a certain escaping probability 
\begin{equation} 
 {\cal P}(x,k)
 =\exp\left[-\int_\tau^\infty\rho\sigma v d\tau'\right], 
 \label{P} 
\end{equation}  
due to the finite dimensions and lifetime of the 
thermalized matter. The integral above is evaluated in 
the proper frame of the particle. Then, the distribution 
function $f(x,k)$ of the expanding system has two 
components, one representing the portion of the fluid 
already free and another corresponding to the part still 
interacting, {\it i.e.}, 
\begin{equation}
f(x,k)=f_{free}(x,k)+f_{int}(x,k)\,. 
\end{equation} 
We may write the free portion as  
\begin{equation}
f_{free}(x,k)={\cal P}f(x,k)\;. 
\label{ffree}
\end{equation}
The integral (\ref{I}) is then rewritten in CEM as 
\begin{eqnarray} 
 I(q,K)
 &=&\int_{\sigma_0}d\sigma_{\mu}K^{\mu}f_{free}(x_0,K) 
 e^{{iqx}} \nonumber \\ 
 &+&\int d^4x\,\partial_\mu[K^{\mu}f_{free}(x,K)] 
 e^{{iqx}},
 \label{Ice} 
\end{eqnarray} 
where the surface term corresponds to particles already 
free at the initial time. 

The problem of this description is its complexity in 
handling because, ${\cal P}$ depends on the momentum of 
the escaping particle and, moreover, on the future of 
the fluid as seen in eq.(\ref{P}). In order to make the 
computation practicable, here we first take ${\cal P}$ 
on the average, {\it i.e.}, 
${\cal P}(x,k)\Rightarrow<{\cal P}(x,k)>
 \equiv{\cal P}(x)\,.$  
Then, approximate linearly the density $\rho(x^\prime)=\alpha s(x^\prime)$ in eq.(\ref{P}).  
Thus, 
\begin{equation} 
{\cal P}(x,k) \Rightarrow {\cal P}(x) 
=\exp\left(-\kappa\frac{s^2}{|ds/d\tau|}\right)\,,
\label{prob} 
\vspace*{-.2cm} 
\end{equation} 
where $\kappa=0.5\,\alpha<\sigma v>$. 

Now, eq.(\ref{Ice}) is translated into SPH language, by 
computing the sum (\ref{corr_sph}) not over $T=T_{fo}$ 
but picking out SPH particles according to this 
probability, with $n_{j\mu}$ pointing to the 
4-gradient of ${\cal P}$. Thus, our approximation 
includes also emission of particles of any momentum, 
once a SPH particle has been chosen. However, since our 
procedure favors emission from fast outgoing SPH 
particles, because $\rho$ decreases faster and so does 
$s$ in this case making ${\cal P}$ larger, we believe 
the main feature of CEM is preserved. 
\smallskip

\ni {\it Results} -- We first assume sudden freezeout 
(FO) at $T_{fo}=128\,$MeV. This temperature was 
previously found by studying the energy dependence of 
kaon slope parameter $T^*$\cite{eff-t}. It has been 
also shown\cite{ebe2} that $T^*$ is not sensitive to IC 
fluctuations. 

\begin{figure}[!htb] 
\vspace*{6.1cm}
\includegraphics{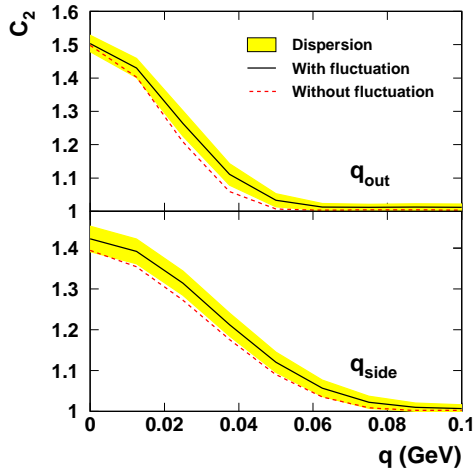}
\caption{\small Correlation functions from fluctuating 
IC and averaged IC. Sudden freeze-out is used here. The 
rapidity range is $-0.5\leq Y\leq 0.5$ and $q_{o,s,l}$ 
which do not appear in the horizontal axis are 
integrated over $0\leq q_{o,s,l}\leq 35$MeV.} 
\label{c2_fic}
\vspace*{-.5cm}
\end{figure}

\begin{figure}[!bth]
\vspace*{10.2cm}
\begin{center}
\includegraphics{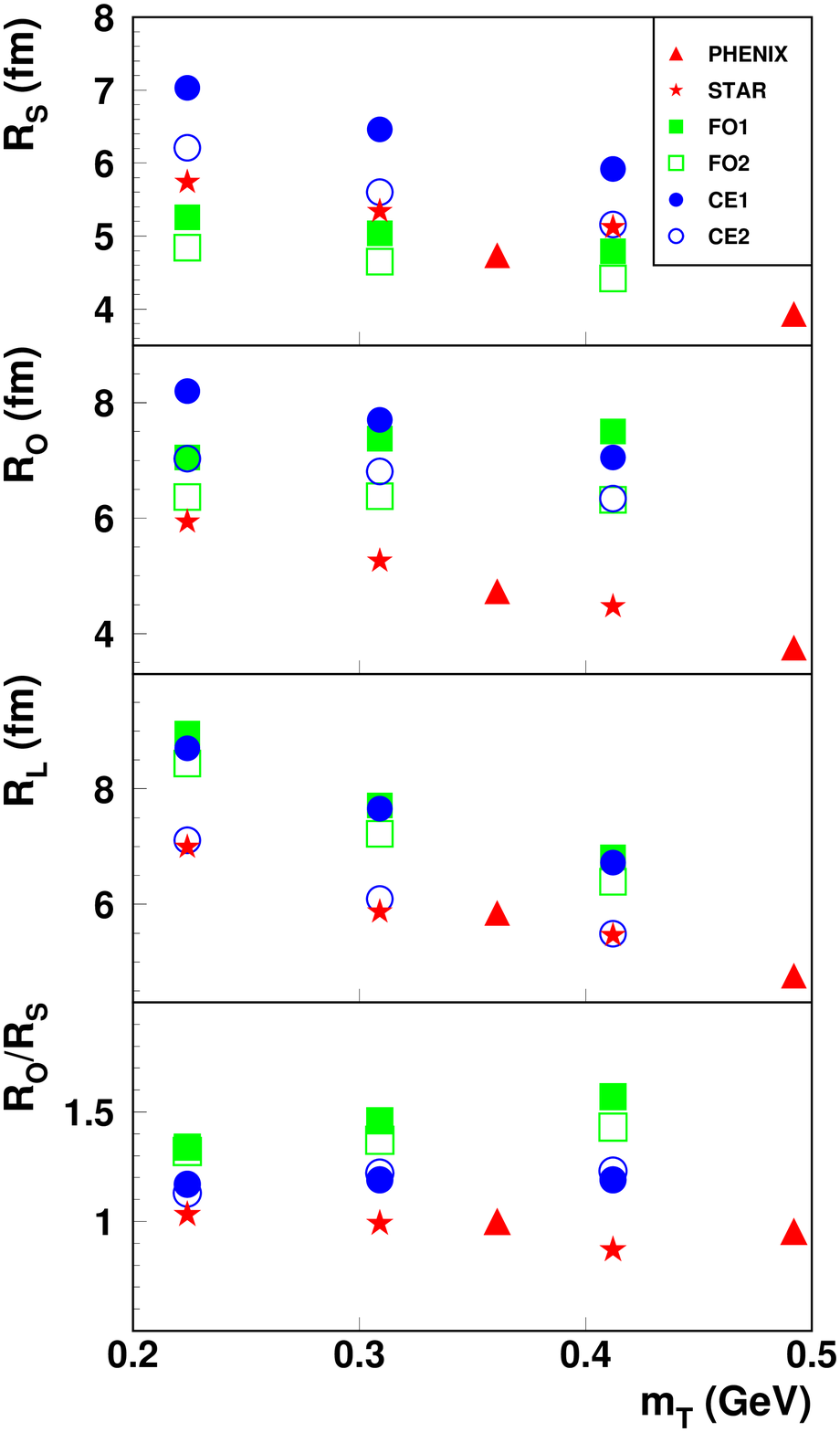}
\end{center}
\vspace{-.9cm} 
\caption{\small HBT radii and the ratio $R_o/R_s$ for 
sudden freeze-out (FO) and CE. 1 stands for averaged IC 
and 2 fluctuating IC. Data are from\cite{star,phenix}: 
$(\pi^++\pi^-)/2$.} 
\label{radii}
\end{figure}

In Fig.\ref{c2_fic}, we compare $C_2$ averaged over 
15 fluctuating events with those computed from the 
averaged IC (so, without fluctuations). One can see 
that the IC fluctuations are reflected in large 
fluctuations also in the HBT correlations. When 
averaged, the resulting $C_2$ are broader than those 
computed with averaged IC, so giving smaller radii as 
expected. Also the shape of $C_2$ changes. We plot the 
$m_T$ dependence of HBT radii, with Gaussian fit of 
$C_2\,$, in Fig.\ref{radii}, together with RHIC data\cite{star,phenix} and\hfilneg\ 

\ni results with CEM. It is seen that the smooth IC 
with sudden FO makes the $m_T$ dependence of $R_o\,$ 
flat or even increasing, in agreement with other hydro 
calculations\cite{morita} but in conflict with the 
data. The fluctuating IC make the radii smaller, 
especially $R_o\,$, without changing the 
$m_T$-dependence. 

\begin{figure}[!htb] 
\vspace*{5.5cm}
\includegraphics{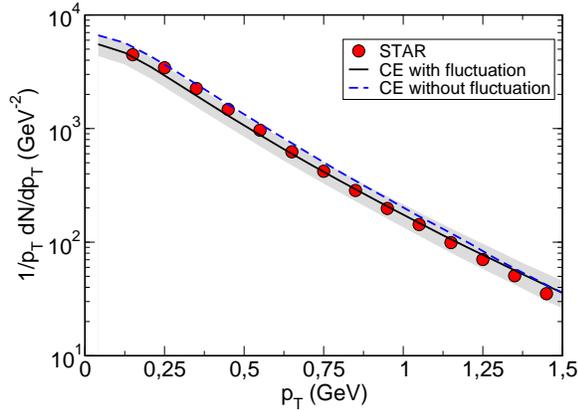} 
\vspace*{-.2cm}
\caption{charged-particle $p_T$ distributions in CEM for 
 the most central Au+Au at 130A GeV. The data are from 
 STAR\cite{star1}.} 
\label{rhic_pt}
\end{figure}

Let us now consider CEM. In this paper, we 
estimated $\kappa$ as being 0.3, corresponding to 
$<\sigma v>=2\,$fm$^2$. In Fig.\ref{rhic_pt}, we show 
the charged $m_T$ distribution to ensure\hfilneg\ 
 
\ni that the estimate above does reproduce correctly 
these data. Now, look at the $m_T$ dependence of the 
HBT radii, shown in Fig.\ref{radii}. Comparing the 
averaged IC case, with CEM (CE1), with the 
corresponding frezeout (FO1), one sees that, while 
$R_L$ remains essentially the same, $R_s$ decreases 
faster and as for $R_o\,$, it decreases now inverting 
its $m_T$ behavior. The account of the fluctuating IC 
in addition (CE2) makes 
all the radii smaller as in FO case, obtaining a nice 
agreement with data for $R_L$ and $R_s\,$, and 
improving considerably the results for $R_o$ with 
respect to the usual hydro description. 

\smallskip
\ni {\it Conclusions and Outlooks} -- In this paper, we 
showed that both the event-by-event fluctuations of the 
IC and the continuous particle emission instead of 
sudden freezeout largely modify the HBT 
correlation of produced pions, so they should be 
included in more precise analyses of data. The IC 
fluctuations give smaller radii, without changing the 
$m_T$ dependence, which is a natural consequence of the 
presence of high-density spots at the early times. 
Continuous particle emission, 
on the other hand, does not change $R_L$ but enhances 
the $m_T$ dependence of $R_s$ and inverts the $m_T$ 
behavior of $R_o\,$, which now decreases with $m_T$ in  
accord with data. This is because, in this description, 
large-$k_T$ particles appear mostly at the early stage 
of the expansion from a thin hot shell of the matter, 
whereas small-$k_T$ particles appear all over the 
expansion, and from larger portion of the fluid. The 
combination of these two effects can give account of 
the $m_T$ dependence of $R_L$ and $R_s$ and improves 
considerably the one for $R_o$ with respect to the 
usual version. 

We shall emphasize that these conclusions could only 
be reached because we have explicitly solved 3+1 
dimensional hydrodynamic equations and not simply 
parametrized the final flow as often 
done\cite{tomasik}. 
The results are preliminary. In applying the CEM we had 
to make a drastic approximation, expressed by 
eq.(\ref{prob}), in order to make it feasible. It is 
likely that this is the reason why the discrepancy in 
$R_o$ still persists. In addition, there exist 
certainly dissipation effects. Since SPH is an 
effective description in terms of parameters 
$\nu_i$ appearing in eq.(\ref{s}), some smoothing 
of short-wave-length Fourier components is  
taken into account through the kernel 
$W({\mathbf x}-{\mathbf x}_{\,i}(t);h)$. 
We preferred not to explicitly include the viscosity at 
this stage, as it is still an open problem of 
hydrodynamics (see more details in Ref.\cite{rv}). 

\smallskip

\ni {\it Acknowledgments} -- This work was partially 
supported by FAPESP (2000/04422-7 and 2001/09861-1), 
CAPES/PROBRAL, CNPq, FAPERJ and PRONEX.


\begin{references} 
\bibitem{landau} L.D. Landau, Izv. Akad. Nauk SSSR 
 {\bf 17}, 51 (1953). 

\bibitem{CF} F. Cooper and G. Frye, Phys. Rev. 
 D{\bf 10}, 186 (1974). 

\bibitem{hbt1} R. Hanbury-Brown and R.Q. Twiss, Phil. 
 Mag. Ser. {\bf 7}, Vol. {\bf 45}, 663 (1954). 

\bibitem{hbt2} C-Y. Wong, {\it Introduction to 
 High-Energy Heavy-Ion Collisions}, World Scientific 
 (1994); 
 R.M. Weiner, {\it Bose-Einstein Correlations in 
 Particle and Nuclear Physics}, J. Wiley \& Sons (1997); 
 U. Heinz and B. V. Jacak, Ann. Rev. Nucl. Part. Sci. 
 {\bf 49}, 529 (1999); 
 T. Cs\"org\H{o}, Heavy Ion Phys. {\bf 15}, 1 (2002). 

\bibitem{hama} Y. Hama and S. Padula, Phys. Rev. 
 D{\bf 37}, 3237 (1988). 

\bibitem{pratt} S. Pratt, Phys. Rev. D{\bf 33}, 1314 
 (1986); 
 K.Kolehmainen and M. Gyulassy, Phys. Lett. B{\bf 180}, 
 203 (1986); 
 A.N. Makhlin and Yu.M. Sinyukov, Z. Phys. C{\bf 39}, 
 69 (1988). 

\bibitem{ebe1} T. Osada, C.E. Aguiar, Y. Hama and 
 T. Kodama, nucl-th/0102011; 
 C.E. Aguiar, Y. Hama, T. Kodama and T. Osada, Nucl. 
 Phys. {\bf A698}, 639c (2002). 

\bibitem{ebe2} Y. Hama, F. Grassi, O. Socolowski Jr., 
 C.E. Aguiar, T. Kodama, L.L.S. Portugal, B.M. Tavares 
 and T. Osada, Proc. of 32$\,$ISMD Symposium, eds. 
 A. Sissakian {\it et al.}, World Scientific 
 (Singapore, 2003), 65. 

\bibitem{hirano} See also T. Hirano, nucl-th/0403042. 

\bibitem{ghk} F. Grassi, Y. Hama and T. Kodama, 
 Phys. Lett. B{\bf 355}, 9 (1995); 
 F. Grassi, Y. Hama and T. Kodama, Z. Phys. {\bf C73}, 
 153 (1996). 
 
\bibitem{laszlo} 
 V. K. Magas et al., 
 Heavy Ion Phys. {\bf 9}, 193 (1999); 
 Phys. Lett. {\bf B459}, 33 (1999); 
 Nucl. Phys. {\bf A661}, 596c (1999); 
 Yu.M. Sinyukov, S.V. Akkelin and Y. Hama, Phys. Rev. 
 Lett. {\bf 89}, 052301 (2002). 

\bibitem{ghks} F. Grassi, Y. Hama, T. Kodama and 
 O. Socolowski Jr., Heavy Ion Phys. {\bf 5}, 417 (1997). 

\bibitem{gs1} F. Grassi and O. Socolowski Jr., Phys. 
 Rev. Lett. {\bf 80}, 1170 (1998); 
 F. Grassi and O. Socolowski Jr., J. Phys. G{\bf 25}, 
 331 (1999).

\bibitem{gs2} F. Grassi and O. Socolowski Jr., J. Phys.  
 G{\bf 25}, 339 (1999). 

\bibitem{hbt-cem} F. Grassi, Y. Hama, S.S. Padula and O. Socolowski Jr., 
Phys. Rev. C{\bf 62}, 044904 (2000). 

\bibitem{star} C. Adler {\it et al.}, Phys. Rev. Lett. 
 {\bf 87}, 082301 (2001). 

\bibitem{phenix} K. Adcox {\it et al.}, Phys. Rev. Lett. 
 {\bf 88}, 192302 (2002). 

\bibitem{nexus} H.J. Drescher, M. Hladik, 
 S. Ostrapchenko, T. Pierog and K. Werner, J. Phys. 
 G{\bf 25}, L91 (1999); Nucl. Phys. {\bf A661}, 604 
 (1999). 

\bibitem{spherio} C.E. Aguiar, T. Kodama, T. Osada and 
 Y. Hama, J. Phys. G{\bf 27}, 75 (2001). 

\bibitem{csorgo} S. Nickerson, T Cs\"org\H o and 
 D. Kiang, Phys. Rev. C{\bf 57}, 3251 (1998). 

\bibitem{eff-t} 
 Y. Hama, F. Grassi, O. Socolowski Jr., T. Kodama, 
 M. Ga\'zdzicki and M.I. Gorenstein, 
 Acta Phys. Polon. {\bf 35}, 179 (2004). 

\bibitem{star1} 
 C. Adler et al., Phys. Rev. Lett. {\bf 87}, 112303 
 (2001). 

\bibitem{morita} K. Morita, S. Muroya, C. Nonaka and 
 T. Hirano, Phys. Rev. C{\bf 66}, 054904 (2002). 

\bibitem{spherio1} {\bf S}moothed {\bf P}article 
 {\bf h}ydrodynamical {\bf e}volution of 
 {\bf R}elativistic heavy {\bf IO}n collisions

\bibitem{tomasik} 
 B. Tom\'a\v{s}ik, U.A. Wiedemann and U. Heinz, 
 Heavy Ion Phys. {\bf 17}, 105 (2003); 
 F. Reti\`ere and A. Lisa, nucl-th/0312024; 
 T. Renk, hep-ph/0404140.  

\bibitem{rv} Y. Hama, T. Kodama and O. Socolowski Jr., 
 hep-ph/0407264. 

\end{references}
\end{document}